# IHEP EXPERIENCE ON CREATION AND OPERATION OF RFQS.

O.K. Belyaev, O.V. Ershov, I.G. Maltsev, V.B. Stepanov, S.A. Strekalovskikh, V.A. Teplyakov,
A.V.Zherebtsov, IHEP, Protvino, Russia

*Abstract*

The new 1.8 MeV proton RFQ was completed and start operation in the IHEP in 1997. It was built according to the plan of modernization of the injection system to the booster of the IHEP proton synchrotron. In this paper we will describe design features of the RFQ and scientific and engineering achievements which was reached since its creation and operation.

## Introduction

An insistent need to modernize the injection system was caused by following reasons. The old injector is under operation since 1983 (the initial part of this linac – RFQ since 1990) [1]. Most of RF and other equipment becomes old and should be modified. Besides it was decided to modify the accelerating system. The attractive feature of this RFQ is a high acceleration rate (1.4 MeV/m at the end (Fig.1)) which allows better matching on longitudinal motion with next part of the linac. It was reached due to proper variation of phase Φ and transit time factor $\vartheta$ along the RFQ (Fig.1) In its turn high values of $\vartheta$ (0.66 at the end) were achieved by using a trapezoidal modulation of electrodes at the end of the RFQ instead of sinusoidal ones (Fig.2).

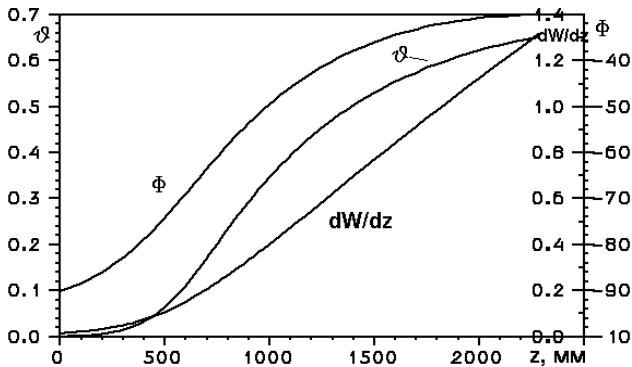

Figure.1: RFQ parameters along the linac

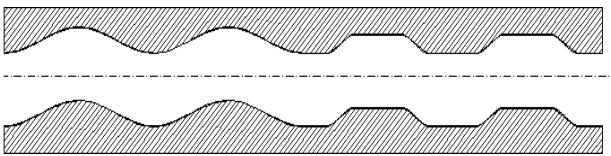

Figure2: Sinusoidal and trapezoidal RFQ electrodes

"Trapezoidal" electrodes allows to obtain high $\vartheta$ without increasing of the modulation factor and hence peak field on the surface [2]. Indeed, the following expression defines $\vartheta$ as a first approximation [3]:

$$\vartheta \approx \frac{\sin(\pi\alpha)}{\pi\alpha}, \qquad (1)$$

where $\alpha = g/\beta\lambda$, g – length of a gap. As this formula shows, $\vartheta$-factor increases with decrease in the gap length. This expression is correct for RFQ structures too. One may define the «gap» here as the space where $E_z \neq 0$, i.e. the whole cell except of a small piece at the top of a sinusoid. But if the sinusoid top is «cut off» then the «gap» will decrease and according to (1) $\vartheta$ increases.

Main parameters of this RFQ are given in the Table.

Table : RFQ parameters

| RFQ parameter | Value |
|---|---|
| Input energy | 0.1 MeV |
| Output energy | 1.8 MeV |
| Operating frequency | 148.5 MHz |
| RF voltage | 140 kV |
| Synchronous phase | -90÷-30 deg |
| Maximum surface E-field | 250 kV/cm |
| Capture | 99% |
| Operating beam current | 120 mA |
| Total length | 2563 mm |
| Characteristic bore radius | 7.59 mm |
| Total number of cells | 91 |
| «sinusoidal» cells | 71 |
| «trapezoidal» cells | 20 |

## Acceleration structure and RFQ option.

The general view of the RFQ is shown in Fig.3.

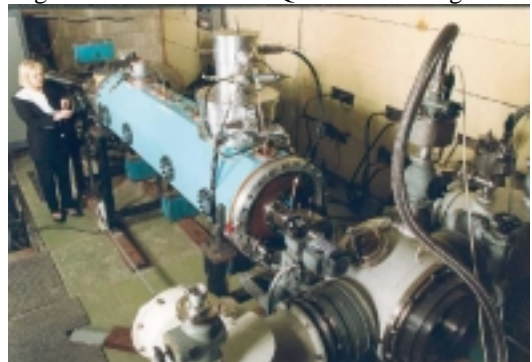

Figure 3: General view of the RFQ.

2H resonator was used as an accelerating structure of the RFQ (Fig.4). It consists of 520 mm diameter RF vac-uum tank and two 164 mm diameter resonators which are pieces of tubes with slits. Accelerating electrodes are mounted at edges of these slits. The tank and resonators are manufactured of aluminium alloy and coated with 100 µm copper layer.

The technology of galvanic coating provides very good polished surfaces with high conductivity. Electrodes are machined of oxygen-free copper. An 50 µm deformable indium seal in the mount unit is used for better electric junction between electrodes and resonators. 3 ion pumps provide a high (of the order $5 \cdot 10^{-8}$ torr) vacuum. Besides additional getter pump can keep a good enough vacuum during downtime of the linac. Water-cooling pumping through channels in the tank and resonators provides required temperature stability.

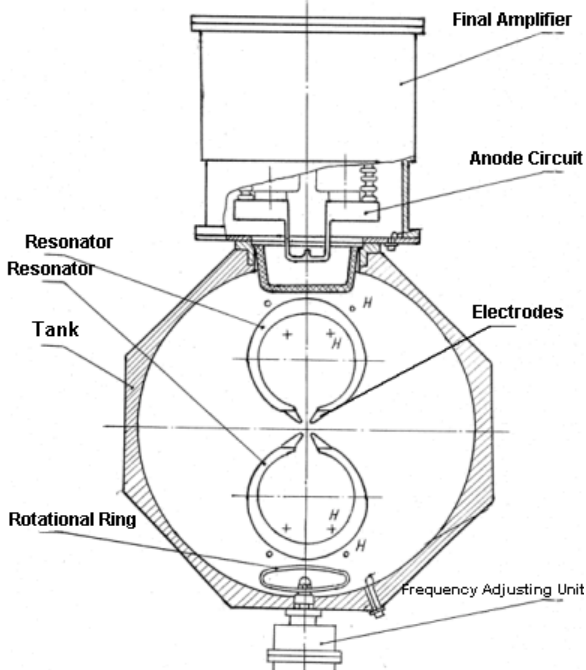

Figure 4: Cross section of the RFQ.

The choice of this accelerating structure was caused due to an absence of parasite oscillation modes in the immediate neighbourhood of the operating mode and technological effectiveness of its manufacture.

RF tuning was carried out by separated action on electric and magnetic fields in front and end faces of the 2H resonator. The small perturbation method was used for measurements of an E-field quadrupole component. Results are shown in the Fig.5. As the figure shows, normalized distribution is well agreed with required curve. The rms deviation is $\sigma = 1.3\%$.

Q-factor for 2H-resonator, calculated with help of PRUD-0 code [4], is equal to 15800. Measured Q-factor is 13600 (86% of calculated one). RF power losses which dissipates in the resonator walls at this value of the Q-factor is about 200 kW.

Tuning of the operating frequency was carried out by rotational rings as adjusting units (Fig.4). Each ring changes a frequency by 8 KHz. High-speed automatic trimming ($\Delta f / \Delta t = 190\ KHz/c$) is used to compensate frequency detuning while the accelerator is under operation. This system also keep a frequency deviation in the range of $\pm 1.5$ KHz.

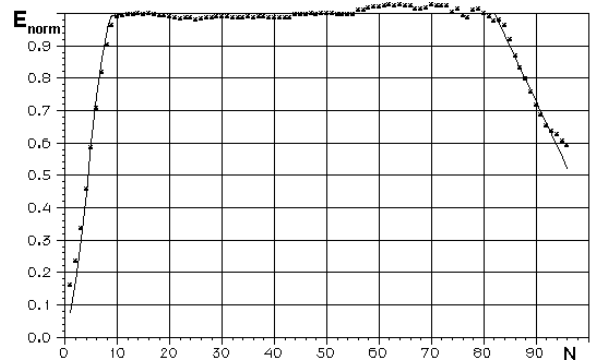

Figure 5: Distribution of the E-field quadupole component as a function of the cell number. The solid curve – requirements, asterisks – measurements.

The regular X-ray radiation control along the RFQ permits to register pollution of electrodes by vapor components from the Ion Source [5]. The surge exhaust volume putting into the LEBT of the RFQ and periodic RF conditioning for cleaning electrodes permitted to increase the break-down limit in the structure up to 450 kV/cm (3.4 $E_k$).

## The RF power system

The RF power is supplied in the resonator with external inductive coupling loop through the vacuum-tight ceramic window placed in the RF tank. It allows to excite the resonator both by the single final power amplifier and power amplifier mounted on the resonator [6].

Fig.6 shows the block scheme of the RF transmission system. The resonator (R) is excited by the two-stage final power amplifier (FA) which is integrated with the RF vacuum tank (Fig.4). The symmetric coupling loop which is an inductive shoulder of the anode circuit links the final cascade (C2) with the resonator. The coupling loop is immersed into the ceramic window (CrW). This link is equivalent to that which is provided via the multiple $\pi/2$ electric length feeder. The C2 was realized using the 12 triodes push-pull amplifier with common greed [7]. The pre-amplifier cascade (C1) is inductively linked with FA and realized according the similar 2-triodes scheme [6]. The pre-exciter (PE) contains quarter-wave anode circuit triodes amplifier cascades (A1), (A2). It is excited by wide-band transistor amplifier (TA). The RF amplitude and phase are controlled by high-speed phase shifter-attenuator (PSA) which is mounted at the TA entry. The operating frequency is generated by 0.5 W stabilized driver (SD). A maximum pulse power of the final amplifier is equal to 600 kW, anode voltage in FA – 12 kV, in

PE − 6 kV. The modulator (M) forms 80 μs pulses of the anode voltage. A method of voltage addition in tiristor cells with pulse transformer help is used in the modulator. A compensation due to the beam load is provided by the PSA and modulator adjusting via a signal from the programmed control system (PCS). Besides two-circuit systems for an amplitude and phase automatic control (ACA and ACP) is under development. To increase the dynamic stabilization under action of the accelerated beam additional adjusting circuit is used which is controlled by the signal from the beam current pick-up (BCP).

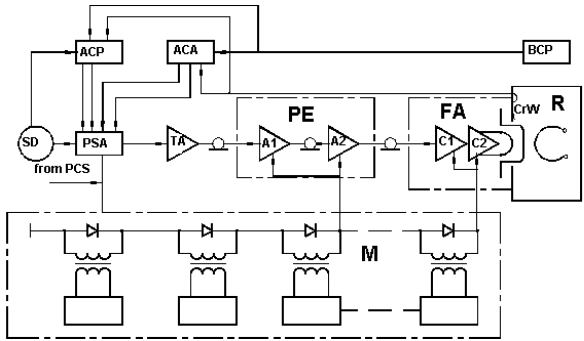

Figure 6: the RF block scheme.

## Beam measurements

Measurements of beam parameters were carried out both in LEBT and HEBT lines. The injection system of the RFQ consists of the Ion Source and LEBT including 2 matching solenoids and heavy masses separating collimators. The total LEBT length is equal to 620 mm. For measurements of the beam parameters the test stand was created. It includes beam diagnostics apparatus and automatic data acquisition system. To obtain a matched beam at the entry of the RFQ a number of measurements of a beam emittance in different points of LEBT are required. Adjusting Ion Source and LEBT parameters including the plasma emitter, two-gap optics and solenoids, permits us to approach required input parameters of the beam. This procedures allows to obtain the output beam emittance, which is 1.5-2 times greater than calculated one. Fig.7 shows measured 120 mA output beam emittance.

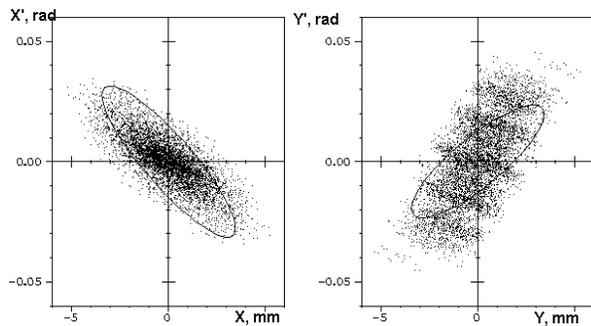

Figure 7: Measured transverse beam emittance. Solid lines – are calculated ellipces.

## Conclusions

The RFQ with permanently growing accelerating rate was created. It allows to obtain better matching with next part of the linac and decrease (1.5 times approximately) the RFQ length.

Materials, technologies of manufacturing and assembly which were used permit to obtain the high vacuum ($5 \cdot 10^{-8}$ torr) and simplify the vacuum system.

Efforts which were made against pollution of electrodes by Ion Source operation products and regular RF conditioning allowed to obtain high value of the breakdown limit (3.4 $E_k$).

The new RF power system with final power amplifier integrated with the RF vacuum tank (Fig.4) was created and tested. New modular concept was used in the high voltage modulator. As a result it demonstrates a reliable 5 Hz operation.